\documentclass[
 reprint,
amsmath,amssymb,
aps,
pra,
]{revtex4-2}

\usepackage{mathtools}
\usepackage{graphicx}
\usepackage{dcolumn}
\usepackage{bm}
\usepackage{multirow} 

\usepackage{braket}

\newcommand{\PT}{$\mathcal PT$}

\usepackage{xcolor}
\usepackage{ulem}

\begin{document}


\title{Effect of imaginary gauge on wave transport in driven-dissipative systems}





\author{I. Komis$^{1,2,3}$} 
\author{K. G. Makris$^{2,3}$}
\author{K. Busch$^{1,4}$}
\author{R. El-Ganainy$^{5}$}

\affiliation{$^1$Humboldt-Universit\"{a}t zu Berlin, Institut f\"{u}r Physik, AG Theoretische Optik \& Photonik, D-12489 Berlin, Germany}
\affiliation{$^2$Institute of Electronic Structure and Laser, Foundation for Research and Technology-Hellas (FORTH), P.O. Box 1527, 71110, Heraklion, Greece}
\affiliation{$^3$ITCP, Department of Physics, University of Crete, 70013, Heraklion, Greece}
\affiliation{$^4$Max-Born-Institut, 12489 Berlin, Germany}
\affiliation{$^5$Department of Electrical and Computer Engineering, Saint Louis University,  Saint Louis, MO 63103, USA}

\date{\today}

\begin{abstract}
Wave transport in disordered media is a fundamental problem with direct implications in condensed matter, materials science, optics, atomic physics, and even biology. The majority of studies are focused on Hermitian systems to understand disorder-induced localization. However, recent studies of non-Hermitian disordered media have revealed unique behaviors, with a universal principle emerging that links the eigenvalue spectrum of the disordered Hamiltonian and its statistics with its transport properties. In this work we show that the situation can be very different in driven-dissipative lattices of cavities, where a uniform gain applied equally to all the components of the system can act as a knob for controlling the wave transport properties without altering the eigenvalue statistics of the underlying Hamiltonian. Our results open a new avenue for developing a deeper insight into the transport properties in disordered media and will aid in building new devices as well. Our work which is presented in the context of optics generalizes to any physical platforms where gain can be implemented. These include acoustics, electronics, and coupled quantum oscillators such as atoms, diamond centers and superconducting qubits.

\end{abstract}


\maketitle

\section{Introduction}

Wave transport in random and disordered media was originally studied in the context of condensed matter physics to explain electron transport properties \cite{Anderson}. The notion of Anderson localization shaped the future of solid-state physics and has been widely investigated both theoretically and experimentally \cite{Gang_Four, Disorder1, Disorder2, DisOpt1, DisOpt2, DisOpt3, DisOpt4}. Over the past decades, it was recognized that disorder plays a central role in shaping wave dynamics in various platforms, ranging from optics/photonics \cite{DisOpt5, DisOpt6, DisOpt7, DisOpt8, DisOpt9, DisOpt10, DisOpt11, DisOpt12, DisOpt13, DisOpt14}, quantum physics \cite{DisEnt1, DisEnt2}, and atomic systems \cite{DisAtSys1, DisAtSys2, DisAtSys3, DisAtSys4} to acoustics \cite{DisAco1, DisAco2, DisAco3} and even biology \cite{DisBio}. Interestingly, it was also shown that effects analogous to wave localization in disordered media are linked to the failure of some quantum algorithms \cite{LQA_PNAS_2020}. With a few exceptions \cite{RandomL1, RandomL2, RandomL3}, a common theme among all studies that considered localization in linear and conservative media, thereby excluding random lasers, is their focus on systems described by Hermitian Hamiltonians having real-valued spectra.

However, recent interest in the physics of non-Hermitian systems and its unique characteristics—such as phase transitions between real and complex eigenvalues in parity-time ($\mathcal{PT}$) symmetric Hamiltonians \cite{Bender1}, the presence of non-Hermitian singularities known as exceptional points (EPs), and the non-orthogonality of eigenvectors \cite{EP1, EP2}—as well as potential applications in optics and photonics \cite{PT2, PT4, PT5, PT6, Appli1, Appli2, Appli3, Appli4}, have led to more careful investigations into wave transport in disordered non-Hermitian media. In this context, the interplay of non-Hermitian disorder and Anderson localization has been recently theoretically investigated. More specifically, among the key studies are the concept of constant intensity waves, where propagation through disordered media is possible without any backscattering \cite{Makris1, Makris2, Makris3, Makris4}, Anderson localization in two- and three-dimensional lattices \cite{Makris5, NHDisorder1}, and transport discontinuities known as ``jumpy" propagation or Anderson jumps in such non-Hermitian disorder lattices \cite{Makris6, Makris7}. These ideas of constant-intensity waves (correlated disorder) and Anderson jumps (uncorrelated disorder) have  been recently demonstrated experimentally in the acoustic domain, as constant-pressure waves \cite{Exp1}, as well as, in the optical regime, as sudden jumps \cite{NHDisorder2} and photonic constant-intensity beams that induce non-Hermitian transparency \cite{Exp2}, based on fiber loop mesh lattices. These developments led to a revived interest in Anderson localization, but from the non-Hermitian perceptive, resulting to investigations of topological effects, including impurities, scaling theories, and topological Anderson transitions \cite{NHDisorder3, NHDisorder4, NHDisorder5, NHDisorder6, Makris8, NHDisorder7}. Additional studies in non-Hermitian quasiperiodic systems, particularly in the non-Hermitian extension of the Aubry-Andre-Harper model, have revealed localization transitions and mobility edges in quasicrystals \cite{AAH, Quasi1, Quasi2, Quasi3, Quasi4, Quasi5}. Along these lines, further experimental studies have examined chaotic disorder \cite{Exp3}, open transmission channels \cite{Exp4}, and Anderson transitions in parity-time symmetric systems \cite{Exp5}.

\begin{figure*}
    \centering
    \includegraphics[width=0.97\textwidth]{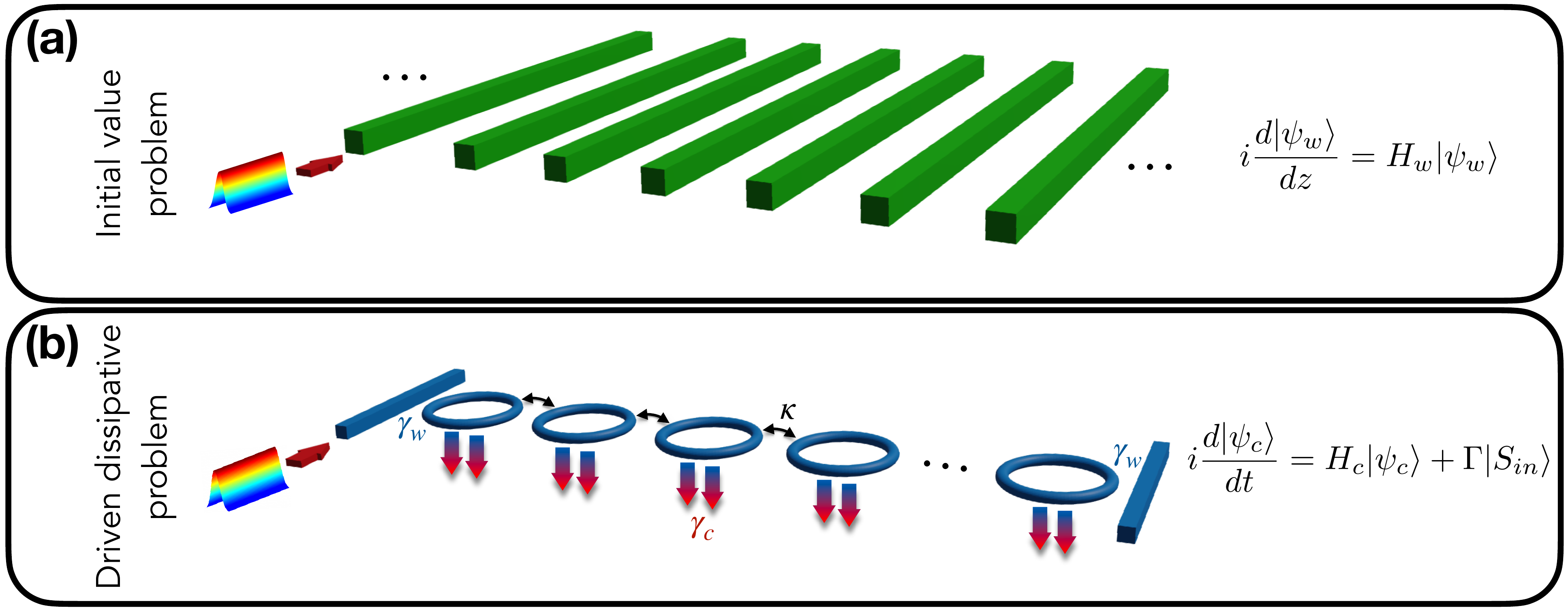}
    \caption{Schematic representation of two distinct photonic systems. (a) An array of coupled waveguides and (b) an array of coupled cavities. The spatial and temporal coupled mode equations governing each system are also shown. Usually, it is tacitly assumed that the spatial coupled mode description of system of coupled waveguides and the temporal coupled mode description of coupled cavities leads to similar results due to the close mathematical resemblance of the corresponding system matrices $H_w$ and $H_c$, respectively. However, the resulting dynamics can be quite different due to the presence of the driving term $\Gamma \ket{S_{in}}$.}
    \label{fig:Schematic}
\end{figure*}

Among the various mathematical techniques used to study wave propagation in random media, a particular organizing principle that plays a central role in classifying different localization and transport schemes is eigenvalue statistics, also commonly known as level spacing statistics \cite{Statistics}. In Hermitian systems with real eigenvalues, they are naturally defined as the spectral gaps between consecutive eigenvalues. The significance of eigenvalue statistics is exemplified by the one-dimensional Anderson model \cite{Anderson}, which is formally known to exhibit localization for any level of disorder \cite{Gang_Four}. For weak disorder, eigenstates overlap spatially, leading to level repulsion, and the eigenvalue statistics resemble the Wigner-Dyson distribution, which predicts that $P(S=0)=0$, where $P(S)$ refers to the probability of the level spacing $S$, a situation analogous to a perfect lattice without disorder, where eigenstates remain delocalized. For strong disorder, eigenstates become localized with smaller localization lengths, minimizing spatial overlap with different states occupying distinct regions across the lattice. Consequently, they can have identical (or near-identical) eigenvalues, i.e., $P(S=0)\neq 0$, and the eigenvalue statistics follow a Poisson distribution, indicating the absence of correlations between different states. The presence or absence of localized states, in turn, affects the transport properties of the system. For example, in uniform waveguide arrays without disorder, light propagation follows the well-known discrete diffraction pattern \cite{DisOpt3}, whereas in disordered arrays, light propagation can exhibit diffusive behavior or even undergo a localization transition \cite{DisOpt4}.

In non-Hermitian systems, the situation is complicated by the fact that eigenvalues are generally complex. As a result, there is no unique rule to organize the spectrum to order the eigenvalues and define eigenvalue statistics. For instance, one may consider the real (or imaginary) parts of the eigenvalues as an ordering criterion. Alternatively, one may consider their absolute value, as has been done in \cite{Makris5,Makris6}. However, to date, these choices are arbitrary and do not arise from an underlying physical principle. This in turn raises the question of whether eigenvalue statistics necessarily governs transport properties in non-Hermitian systems, as it does in their Hermitian counterparts.

In this work, we demonstrate that this widely held expectation does not always hold. Specifically, we focus on driven non-Hermitian linear systems operating below the lasing threshold. Such systems are described by non-Hermitian Hamiltonians even in the absence of coupling to input and output channels. Instead of the standard initial value problem commonly studied in disordered systems, we adopt an input-output formalism for our analysis. We show that in this setup, wave transport can be independent of the level spacing properties associated with the underlying non-Hermitian Hamiltonian of the system. These results open new directions in understanding and tailoring wave dynamics in disordered and random media. While our analysis is generic and applicable to any discrete disordered system described by a tight-binding Hamiltonian, we focus on photonic setups due to their well-established role in exploring exotic features of non-Hermitian physics. However, our results also apply to discrete atomic, electronic, and acoustic setups.


\section{Physical model of coupled cavities}
Discrete non-Hermitian optical systems have become a focal point of theoretical and experimental research, with implementations of these systems frequently utilizing waveguide or cavity arrays, as shown in Fig.~\ref{fig:Schematic}(a) and (b). Within the framework of coupled mode analysis, these systems can be described by the spatial and temporal coupled mode theories (CMT), respectively:

\begin{align}
   i \frac{d\ket{\psi_w}}{dz} &= H_w \ket{\psi_w} \\
   i \frac{d\ket{\psi_c}}{dt} &= H_c \ket{\psi_c} +\Gamma \ket{S_{in}} \label{eq:CMT}
\end{align}
In the above, $\ket{\psi_{w,c}}$ represent the field amplitude vectors in the waveguide or cavity array, respectively and $H_{w,c}$ are the corresponding Hamiltonians. These Hamiltonians describe an ideal one-dimensional system with nearest-neighbor coupling and in the one-band tight-binding context they are tridiagonal, encoding information about the individual elements (propagation constants or resonant frequencies in the diagonal) as well as their interactions (coupling coefficients in the off-diagonals). Moreover, $z$ is the propagation distance along the waveguide array and $t$ denotes time.

Mathematically, our Hamiltonians can be described by, 

\begin{equation}
    H_{w,c} = \kappa \sum _{n=1} ^{N-1} \big( \ket{n}\bra{n+1} + \ket{n+1}\bra{n} \big) + \tilde{\omega}\sum _{n=1} ^{N} \ket{n}\bra{n}
\end{equation}
where $n$ is the index of the elements with $n \in {1, 2, \dots N}$, $\kappa$ denotes the coupling between the elements and $\tilde{\omega}$ are defined as, $\tilde{\omega} \equiv \omega_0 - i\gamma_c -i\gamma_w(\delta_{n,1}+\delta_{n,N})$ (in the case of waveguides it is replaced with their propagation constants $\beta$). Additionally, in the cavity array, when the input and output ports are considered, we have two additional terms entering as loss through the waveguides $(\gamma_w)$. In our setups we consider identical cavities with the same resonant frequency and cavity losses. Finally, $\Gamma$ is the coupling matrix between the array elements and the external excitation channels and $\ket{S}$ is the excitation vector. Additionally, in the case of cavity arrays, Eq. \ref{eq:CMT} is supplemented by the relation, $\ket{S_{out}} = \ket{S_{in}} - i\Gamma^{\dagger} \ket{\psi_c}$ that connects the internal fields $\ket{\psi_c}$ to any relevant output $(\ket{S_{out}})$ and input $(\ket{S_{in}})$ channels \cite{SFan}.

Given the mathematical analogy between spatial and temporal coupled mode theories, it is tacitly assumed that these two systems are equivalent. Consequently, the results obtained for one platform are automatically mapped on to the other. This mapping however, ignores a fundamental difference between the two setups, namely that the spatial CMT is an initial value problem whereas the temporal CMT describes a driven-dissipative system (see Fig. \ref{fig:Schematic}). Thus, while the two setups share the same spectral features for identical arrays (i.e. when $H_w=H_c)$, there is no reason to believe that they must exhibit identical wave transport dynamics. As we demonstrate in this work, non-Hermitian effects can indeed lead to very different behavior of these two systems even when they have identical eigenvalue statistics. In what follows, we focus only on cavity arrays.

Before we continue, we comment here on the choice of the values of the parameters. In this work, we consider arbitrary units, normalizing the parameters of the coupled resonator optical waveguide (CROW) system for simplicity and generality. The resonant frequency is considered to be $\omega_0 = 0$ and the coupling constant between the cavities is set to $\kappa = 1$, representing typical coupling in such systems. The inherent cavity losses are assigned a value of $\gamma_c = 0.1$, while the loss parameter due to the coupling to the waveguides is set to $\gamma_w = 2$, reflecting that the waveguide coupling generally introduces larger losses compared to the inherent cavity losses. As a result, the coupling to the waveguides is higher than the coupling between the cavities, indicating stronger interaction with the external ports than between the cavities themselves. We emphasize that the values of the above parameters are chosen for illustration purposes, without any loss of generality or physical relevance. These normalized values correspond to actual physical conditions in CROW systems. 

\begin{figure*}
    \centering
    \includegraphics[width=1\textwidth]{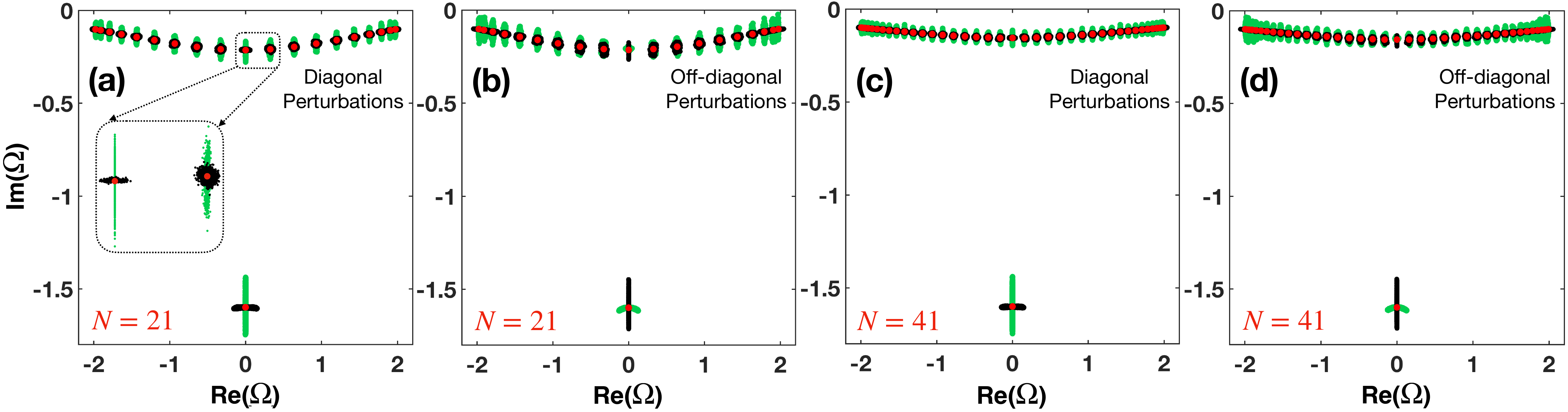}
    \caption{Spectra (red dots) and structured pseudospectra (black and green dots) of our systems. (a), (b) The array consists of $N=21$ cavities with diagonal and off-diagonal perturbations, respectively. Black dots represent real perturbations, while green dots correspond to imaginary perturbations, both with strength $\varepsilon = 0.1$. We consider $l=1000$ different realizations. (c), (d) Similar results for $N=41$ cavities. Notably, the perturbations produce contrasting effects near the central eigenvalues (Re$(\Omega) = 0$), where a sharp line emerges, as shown in the inset of (a). This behavior directly follows from particle hole symmetry and for diagonal perturbations, the sharp line appears when the perturbations are imaginary, whereas for off-diagonal ones, it forms when the perturbations are real.}
    \label{fig:Spectra_PHS}
\end{figure*}

\section{Eigenspectra analysis of the periodic arrays}

We start our analysis by first characterizing the spectral features associated with a passive (i.e. without any gain) discrete one dimensional non-Hermitian array similar to that depicted in Fig. \ref{fig:Schematic}(b) under various types of Hermitian and non-Hermitian perturbations. To do so, we use a computational framework based on pseudospectra \cite{TrefethenBook}. The most basic definition of the $\varepsilon$-pseudospectrum of a non-Hermitian matrix $H_c$ , with $\sigma(H_c)$-spectrum, is the union of all spectra of the matrices $H_c + Z$, where $Z$ is a complex random matrix, with $||Z|| < \varepsilon$, where $||..||$ is the matrix norm, defined by $||A||=\underset{x\neq 0}{\sup}\frac{||Ax||}{||x||}$ \cite{TrefethenBook, Trefethen1, Trefethen2, Trefethen3, pseudo1}. In other words, we study the spectrum ($\Omega$) of the perturbed Hamiltonian $H_c + Z$ for many different realizations $(l)$ of the matrix $Z$. Importantly, to ensure that the comparison between different realizations is meaningful, the matrix $Z$ is normalized according to:
\begin{equation}
   Z = \varepsilon \frac{\mbox{Z}}{\left\Vert \mbox{Z} \right\Vert}
   \label{normalization}
\end{equation}
where $\varepsilon$ indicates the perturbation strength and $\text{Z}$ is the perturbation matrix prior to normalization. At this point we note that, when the perturbation matrix $\text{Z}$ has a particular structure, then we refer to structured pseudospectra \cite{TrefethenBook, Perturbations}, in contrast to complex pseudospectra, where $\text{Z}$ is a full complex random matrix. Here we focus our attention on structured pseudospectra only, meaning diagonal ($\mbox{Z}_{nm} = \delta_{n,m}\cdot z_n$) and off-diagonal ($\mbox{Z}_{nm} =\delta_{n+1,m} \cdot z_n + \delta_{n,m+1} \cdot z_m$) perturbations where the matrix elements of $\text{Z}$ ($z_{nm}$) are drawn from a normal distribution with zero mean and standard deviation equal to one. Notably, our results are independent of the specific choice of distribution, such as uniform instead of normal. In the case of diagonal perturbations, the non-zero elements of matrix $\text{Z}$ correspond to deviations in the resonant frequencies of individual cavities, whereas, in off-diagonal ones, they represent perturbations in the couplings between the cavities. We investigate both real ($z_{nm} \in \mathbb{R}$) and imaginary ($z_{nm} \in i\mathbb{R}$) perturbation scenarios, or equivalently Hermitian and non-Hermitian scenarios, respectively, to fully explore the system's behavior. 

Figures \ref{fig:Spectra_PHS}(a) and (b) depict the distribution of complex eigenvalues ($\Omega$) for an array made of $N=21$ sites for diagonal and coupling perturbations, respectively, for both real (black dots) and purely imaginary (green dots) perturbations. In all cases, the perturbation strength is $\varepsilon=0.1$. Similarly, in Figures \ref{fig:Spectra_PHS}(c) and (d), but for an array of $N=41$ sites. In all subplots, red dots represent the spectrum of the ideal system and since, the number of sites was taken to be odd, there is always a zero mode in the absence of perturbations \cite{PHS1, ZeroMode1, ZeroMode2}. From Figs.~\ref{fig:Spectra_PHS}(a),(c), we observe that for real perturbations (black dots), the eigenvalues spread over a cloud-like structure in the complex domain. A similar behavior is observed for imaginary perturbations except for the central eigenvalue (the eigenvalue of the zero mode), where the distribution under such perturbations spreads over a sharp line on the imaginary axis. Close inspection shows that using only one non-zero element along the diagonal of the matrix $Z$ actually produces an eigenvalue perturbation along a line in the complex domain (see also Appendix A). Each diagonal element, however, corresponds to a line of different slope. Together, these different lines form the cloud. However, in the case of imaginary perturbations, all the diagonal elements have the same slope leading to the creation of the line. The opposite behaviour can be observed for the coupling perturbations (Figs.~\ref{fig:Spectra_PHS}(b),(d)) where real perturbations (black dots) form the sharp line on the imaginary axis. 

The occurrence of the eigenvalues in a line with $\text{Re}(\Omega)$ for real off-diagonal perturbations in Figs.~\ref{fig:Spectra_PHS}(b),(d) is a result of the chiral symmetry of the system. On the other hand, the sharp line on the imaginary axis related to the imaginary diagonal perturbations of Figs.~\ref{fig:Spectra_PHS}(a),(c) can be explained by the particle-hole-symmetry (also known as charge conjugation symmetry) associated with the system under consideration. While this terminology originates from condensed matter physics, where it describes the relationship between electron and hole states in tight-binding models, here we use its mathematical formulation in the photonic context. More specifically, we note that the Hamiltonian matrix $H$ associated with the bipartite lattice structure of Fig. \ref{fig:Schematic} can be expressed in different bases that clusters the two different sublattice groups (even and odd sites), i.e. in the form

\begin{equation}
H_{eo} =  \begin{bmatrix}
        iZ_o & K \\
        K & iZ_e
    \end{bmatrix},
\end{equation}
where the block diagonal matrices $Z_{o,e}$ contain the imaginary perturbations at the odd and even sites, respectively, as defined in Eq.~\ref{normalization}. The coupling between neighboring odd and even sites is represented by the block diagonal matrix $K = \kappa I$, where $I$ is the identity matrix of appropriate dimension. It is straightforward to check that $\chi^{-1} H \chi = -H^*$, with the block diagonal chiral operator $\chi =  \begin{bmatrix}
        1 & \\
          & -1
    \end{bmatrix}$ \cite{PHS1, ZeroMode1, ZeroMode2}. As a side note, we remark that the above relation can be also cast in the form $C^{-1}HC=H$ where the anti-unitary charge conjugation operator $C$ is given by $C=\chi \mathcal{K}$ with the conjugation operator $\mathcal{K}$ \cite{PHS2}. This relation shows that the eigenvalues of $H$ must be symmetrically distributed around the imaginary axis. This in turn implies one of two possibilities: either the eigenvalues are paired symmetrically around the imaginary axis or they lie on the imaginary axis. Remember that, in the example we considered, the number of sites was taken to be odd which implies the presence of a zero mode. Given that small perturbations can only change the eigenvalues by a small amount, this zero mode will remain isolated even after introducing the perturbation and will thus be pinned by symmetry to change only along the imaginary axis. We further explore this for small systems in appendix A.


\section{Spectral Analysis of the Disordered Arrays}
We now turn our attention to the spectral properties and the corresponding eigenvalue statistics of the disordered arrays. Disorder is introduced either in the resonant frequencies, resulting in a total Hamiltonian:

\begin{equation}
    H_c = \sum _{n=1} ^{N-1} \left(\ket{n}\bra{n+1} + \ket{n+1}\bra{n}\right)  + \sum _{n=1} ^{N} (\tilde{\omega}_n + \epsilon_n) \ket{n}\bra{n}
\end{equation}
or in the coupling coefficients, expressed as:
\begin{align}
    H_c = &\sum _{n=1} ^{N-1} \big( (1+\epsilon_{n,n+1}) \ket{n}\bra{n+1} + (1+\epsilon_{n+1,n}) \ket{n+1}\bra{n} \big) \nonumber \\
    &+ \sum _{n=1} ^{N} \tilde{\omega}_n \ket{n}\bra{n}
\end{align}
In both cases, the strength of disorder is characterized by the random variable $\epsilon_n$, that takes random values drawn from a rectangular distribution as,

\begin{equation}
    \text{Re}(\epsilon_n) \in \big[ -\frac{W_R}{2}, \frac{W_R}{2} \big], \;\;\;
    \text{Im}(\epsilon_n) \in \big[ -\frac{W_I}{2}, \frac{W_I}{2} \big],
\end{equation}
where $W_R$ and $W_I$ denote the strengths of the real and imaginary components of the disorder, respectively. For simplicity, we assume these strengths are equal $(W_R = W_I = W)$. Two levels of disorder strength are examined: $W = 0.2$ and $W = 1$, with results averaged over 5000 disorder realizations.

\begin{figure}
    \centering
    \includegraphics[width=0.47\textwidth]{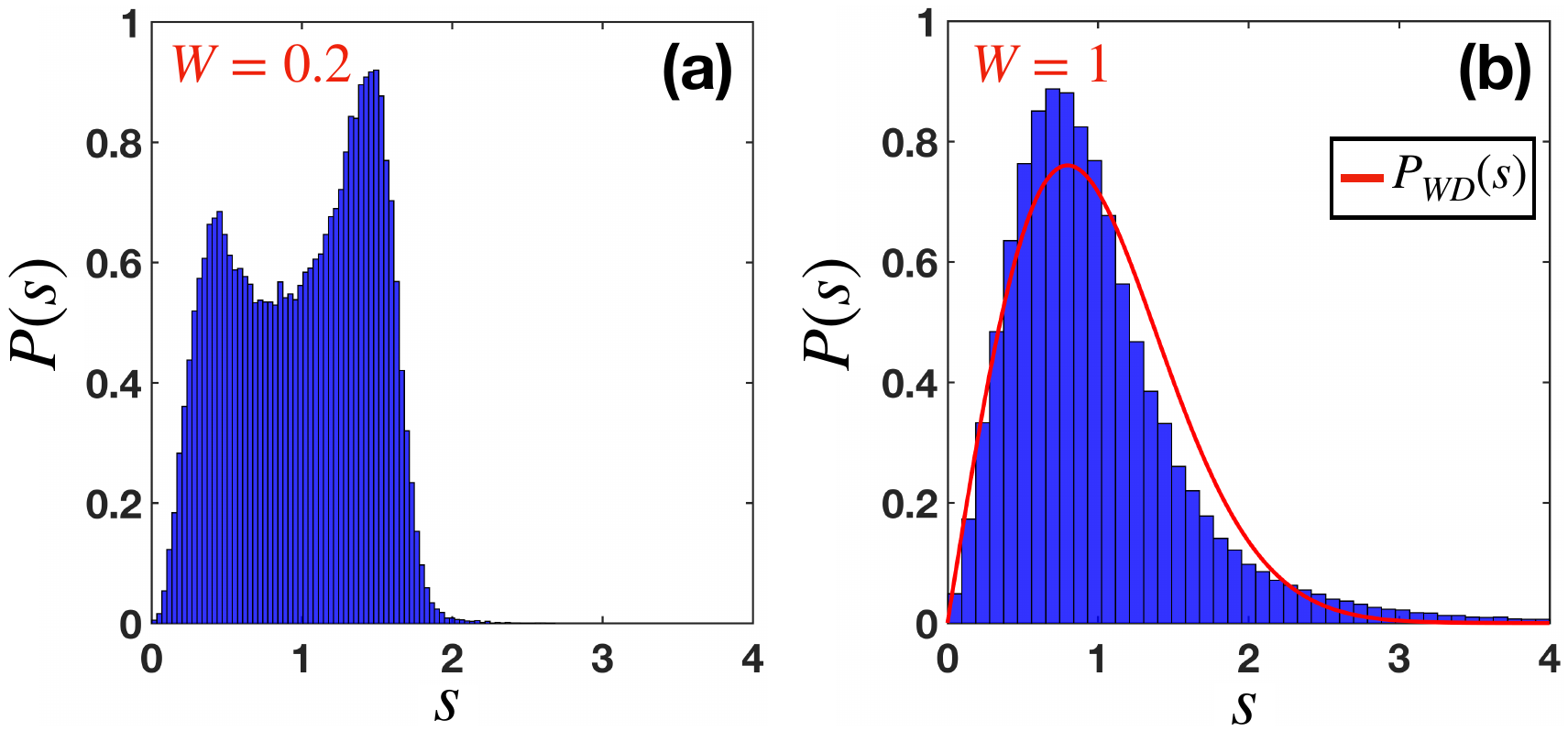}
    \caption{(a), (b) Eigenvalue statistics for 5000 realizations of coupling disorder for two different strengths: $W=0.2$ and $W=1$, respectively. Red line in (b) represents the Wigner-Dyson distribution. The statistical behavior of the eigenvalues remains the same for other types of disorder.}
    \label{fig:Dis_Spec}
\end{figure}

To quantify the eigenvalue statistics, $P(s)$, we follow the criterion outlined in \cite{Makris5}. Specifically, we compute the normalized minimum distance between two eigenvalues in the complex plane, defined as $s = \min{\big| {\Omega_j - \Omega_{j^\prime} }\big|}$. Our results are shown in Fig.~\ref{fig:Dis_Spec} for the case of coupling disorder. In Fig.~\ref{fig:Dis_Spec}(b), the red line represents the Wigner-Dyson distribution, $P_{WD}(s) = \frac{\pi s}{2}\exp{(-\pi s^2 /4)}$ \cite{Statistics}. 

It is important to note that for weak disorder $(W = 0.2)$, the eigenvalue statistics deviates from the Wigner-Dyson form. This behavior is attributed to the relatively small system size $(N = 41)$, which is insufficient to capture the universal statistical properties of random matrix ensembles. In this regime, the eigenstates remain delocalized, and finite-size effects dominate the eigenvalue statistics. For weak disorder $(W < 0.5)$, larger system sizes $(N > 100)$ are typically required for the eigenvalue statistics to converge to the Wigner-Dyson form. In contrast, for stronger disorder $(W = 1)$, the eigenvalue statistics exhibit the expected Wigner-Dyson distribution even for $N = 41$, because the increased disorder strength induces significant eigenstate localization and suppresses finite-size effects. Additionally, for disorder strengths exceeding $W>1$, the eigenvalue statistics retains its Wigner-Dyson form, while the spectrum extends further into the complex plane.


\section{Wave Transport in Driven-Disordered Arrays}
In this section, we investigate the wave transport in the driven-dissipative array described in Fig.~\ref{fig:Schematic}(b). Here, however, we consider passive and active linear arrays that operate below the lasing threshold. The latter are simply the same as the passive arrays, but with an additional uniform gain $g$, meaning the same identical gain value is applied to every cavity in the array as described in \cite{Uniform}. The total Hamiltonian for such systems is given by, 

\begin{equation}
    H = H_c + ig\sum _{n=1} ^{N} \ket{n}\bra{n}
\end{equation}
The parameter $g$ represents the uniform gain added to the system. In general, this gain can be combined with the inherent losses of the cavities $(\gamma_c)$ resulting in a total effective shift to the spectrum of the system. When $g$ equals zero, the system corresponds to the passive array discussed up to this point. If $g$ is negative, it effectively increases the cavity losses, as the term can be absorbed into $\gamma_c$. This framework allows $g$ to be treated as a uniform parameter that is the same for each cavity, representing either gain $(g>0)$ or increased loss $(g<0)$.

\begin{figure}
    \centering
    \includegraphics[width=0.47\textwidth]{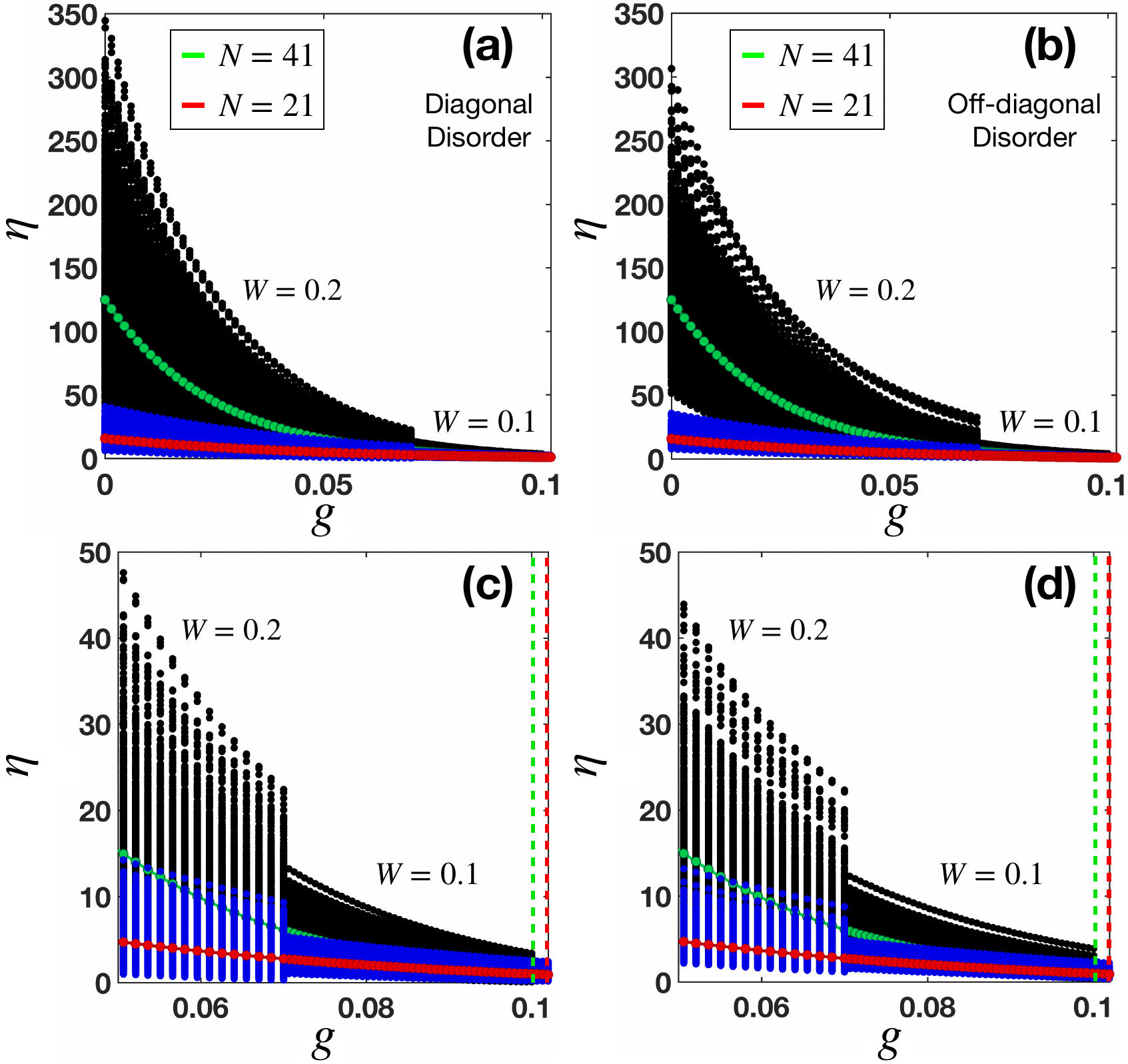}
    \caption{Wave transport in disordered cavity arrays with $N=21$ (red line, blue dots) and $N=41$ (green line, black dots) cavities under different types of disorder: (a) Non-Hermitian diagonal disorder and (b) non-Hermitian off-diagonal disorder. In both cases the disorder strength is $W = 0.2$ for $g \leq 0.07$ and then reduced to $W=0.1$ for $0.07< g < 0.102$. (c), (d) Zoom in of (a) and (b), respectively, highlighting the effects of different disorder strengths. The green dashed line indicate the lasing threshold for $N=41$ cavities, while the red dashed line indicates the lasing threshold for $N=21$ cavities.}
    \label{fig:Transport}
\end{figure}

A critical observation here is that, regardless of how complex eigenvalue statistics are defined, this uniform gain does not alter this statistics, since it introduces a constant imaginary shift to the entire diagonal of the matrix $H$ and hence a constant and identical imaginary shift to all eigenvalues. Intuitively, one might expect that such a trivial shift (which is an imaginary gauge) would not significantly influence the wave transport dynamics, since in the case of waveguides we can easily express the field amplitude at any distance $z$, as $| \psi_w(z) \rangle = e^{g z} e^{-i H_w z} | \psi_w(0) \rangle$. However, our results indicate that this is not correct in a driven-dissipative array. In Fig. \ref{fig:Transport}, we plot the ratio between the power at the first and last elements of the array, namely the transport $\eta \equiv \big| \ket{\psi_{c,1}}/\ket{\psi_{c,N}} \big|^2$ (inverse of transmission), as a function of the applied uniform gain $g$. We consider both the periodic and the disordered arrays while ensuring that the system remains below the lasing threshold. The array is excited from the left waveguide with $\ket{S_{in}} = \ket{S_0}e^{-i\omega_0 t}$ (remember though that in the numerical simulations we take $\omega_0=0$, which is a trivial shift to a rotating frame that does not affect the power calculations). In performing these simulations, we used the same numerical values of the array parameter as before. Hence, the coupling between the edge cavities and the waveguides is $\sqrt{2 \gamma_w}=2$.

\begin{figure}
    \centering
    \includegraphics[width=0.47\textwidth]{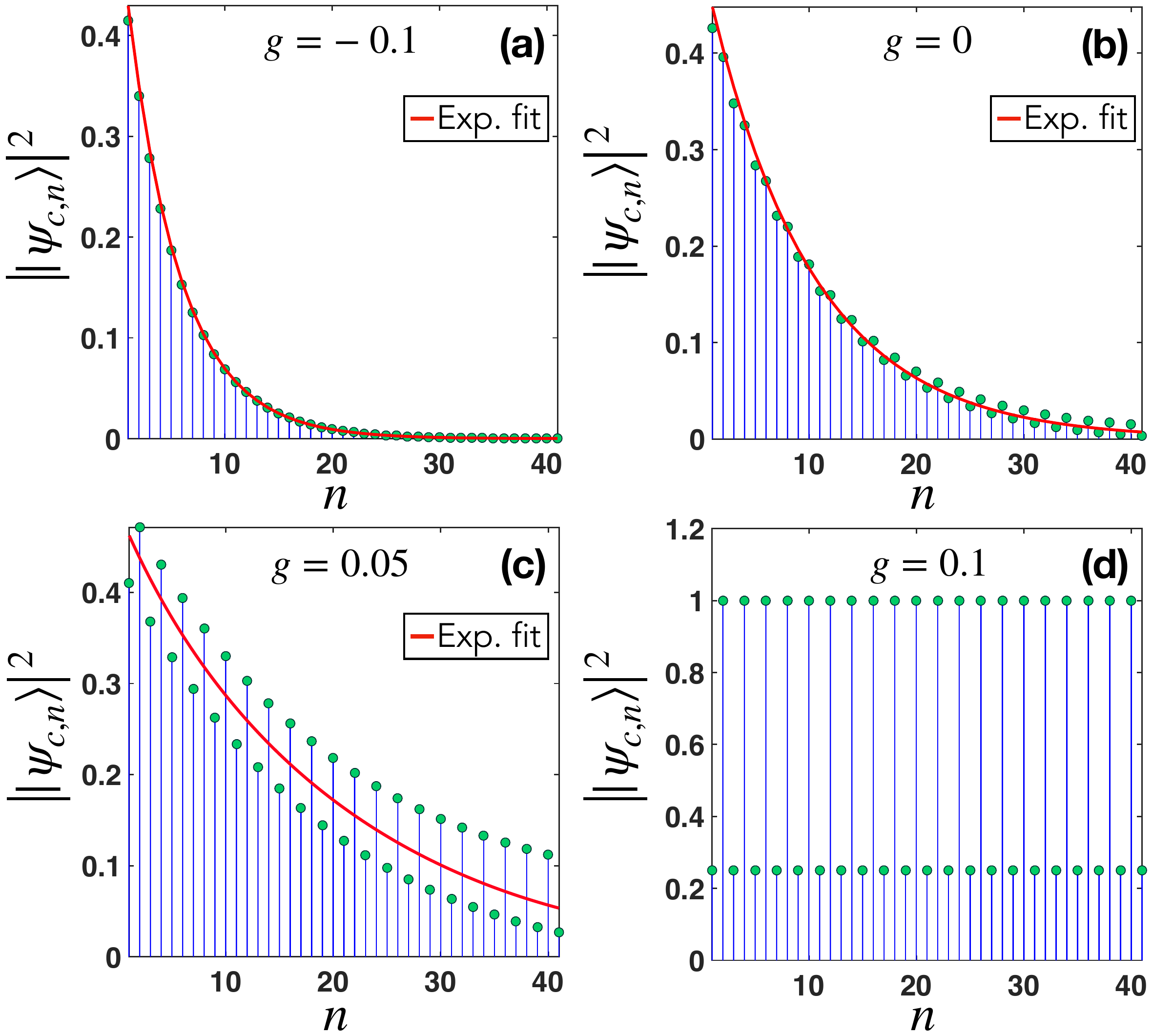}
    \caption{Intensity distribution across each cavity for a system with $N=41$ cavities under varying levels of uniform gain, $g$ and no disorder. (a) $g=-0.1$ corresponding to higher cavity losses, (b) $g=0$, (c) $g=0.05$, and (d) $g=0.1$. Note that for small values of $g$ the amplitudes exhibit an exponential decay (red line). As the uniform gain increases, the behavior changes significantly and the decay is no longer exponential (c).}
    \label{fig:Power}
\end{figure}

Figures \ref{fig:Transport}(a) and (b) depict the transport behavior as a function of $g$ for systems with diagonal and off-diagonal disorders, respectively, with a disorder strength of $W=0.2$ up to $g=0.07$ and $W=0.1$ for $0.07<g<0.102$. This adjustment is necessary because increasing $g$ brings the system closer to its lasing threshold, which occurs at $g \approx 0.102$ for $N=21$ and at $g \approx 0.1002$ for $N=41$. To ensure that the disordered arrays remain below the lasing threshold, the disorder strength must be sufficiently low. In both cases, the red line represents the transport behavior of a system of size $N=21$ in the absence of disorder, while the green line corresponds to a disorder-free system with $N=41$ cavities. The blue and black dots indicate the results of $l=1000$ disorder realizations, where blue corresponds to a lattice size of $N=21$ and black to $N=41$. From these plots, we observe that the transport behavior exhibits a nontrivial dependence on the applied uniform gain $g$. Additionally, in the passive system $(g=0)$, the transport $\eta > 1$, signifies that the power in the first cavity exceeds that in the last. This is expected since our system is excited from the left port adjacent to the first cavity. As $g$ increases, $\eta$ decreases monotonically until it reaches unity, where the power in the last cavity becomes equal to that in the first cavity, adjacent to the excitation waveguide. Interestingly, for certain disorder realizations, it is clear that $\eta <1$ indicating that more power is in the last cavity, an effect achieved solely by tuning the uniform gain. It is important to note that for higher disorder strengths, the range of accessible uniform gain values is limited, as the system reaches the lasing threshold. Nevertheless, we observe the same overall transport behavior, confirming that transport $\eta$ can be controlled by adjusting the uniform gain.

While the plots in Fig.~\ref{fig:Transport} clearly demonstrate that the transport dynamics in these driven non-Hermitian arrangements can vary significantly for systems with the same eigenvalue statistics, they do not fully capture the steady-state optical power distribution in the array under the prescribed excitation scheme. To address this, we plot the intensity distribution inside the array for different values of the uniform gain $g$, in Fig.~\ref{fig:Power}. For a passive linear array and relatively moderate gain values, the steady-state optical power predominantly localizes near the excitation site, as shown in Fig.~\ref{fig:Power}(a) and Fig.~\ref{fig:Power}(b). Note that in Fig.~\ref{fig:Power}(a), $g = -0.1$ indicates that the array experiences greater loss than in the case of $g = 0$. This behavior is consistent with systems experiencing minimal gain, where energy remains concentrated around the initial excitation point due to limited coupling efficiency. Notably, these intensity distributions align well with exponential functions, highlighted by the red line, emphasizing the localized nature of energy propagation at these gain levels. Near the lasing threshold ($g \approx 0.1002$) the intensity distribution undergoes a transition, becoming evenly distributed across the odd and even array elements. At this point, gain plays a dominant role in determining the transport dynamics. When considering the potential impact of disorder on these distributions, small values of $g$ result in behavior that remains largely unchanged, with the exponential fit still holding. However, at higher gain values, disorder can cause notable variations, with some realizations showing increased power in specific regions of the array. In particular, disordered lattices may, in certain cases, transition into the lasing regime, which could explain these deviations. For low levels of disorder, carefully chosen to avoid the lasing regime, the distribution changes slightly but remains broadly consistent. Instead of a perfectly uniform profile, subtle enhancements may appear either at the edges or near the center of the array. These results clearly demonstrate that the transport dynamics in driven non-Hermitian systems can be dramatically altered without varying the eigenvalue statistics. This rather surprising conclusion shed more light on the intriguing behavior of non-Hermitian random media and may open the door to devising new methods for controlling light transport in disordered structures.


\section{Conclusions}
In this work, we have explored the impact of disorder and gain on wave transport and intensity distribution in a driven-dissipative coupled optical resonator array. We have first analyzed the spectral characteristics of the array under different types of disorder, presenting a theoretical framework to explain some of the observed spectral features. Next, we have investigated the wave transport dynamics and optical power distribution across and inside the array under a steady-state excitation, and have demonstrated that both features can vary widely as a function of the uniform gain factor. Remarkably, our results have shown that transport dynamics can vary significantly even for arrays with identical eigenvalue statistics. Additionally, we have found that the optical power distribution, when side-excited, has been localized in systems operating below the lasing threshold but has become completely spread across the array as the system has approached the lasing threshold.

The observed exponential power decay and its subsequent transition to a uniform distribution with increasing uniform gain have borne notable parallels to behaviors in topological systems. For instance, in \cite{NH_Topology, NH_Disorder}, it has been demonstrated that directional amplification in non-Hermitian topological systems has been characterized by exponential gain scaling with system size, governed by a non-trivial winding number of the dynamic matrix. Although our system lacks topological invariants, the exponential power decay along the resonators at low gain levels has closely resembled the suppression of reverse gain reported in these works. This has suggested that similar mathematical principles might govern these dynamics, even in the absence of topological protection. Moreover, at high gain levels, we have observed a transition to a nearly uniform power distribution along the resonator array. This behavior can be compared to the uniform amplification achieved in topologically non-trivial regimes, where topology ensures robustness against disorder and system imperfections. In contrast, our system has relied purely on the interplay between uniform gain and coupling dynamics, making it potentially more sensitive to parameter variations or perturbations.

These comparisons have highlighted the versatility of non-topological systems in exploring gain-loss phenomena while also emphasizing the robustness and control afforded by topological designs. These results not only have challenged some of the conventional wisdom in the field of wave physics in random media, but also have provided more insight into the interaction between disorder and non-Hermiticity in the context of wave propagation, which may lead to developing new strategies for controlling light transport and trapping.


\begin{acknowledgments}
This project was funded by the European Research Council (ERC-Consolidator) under the grant agreement No. 101045135 (Beyond Anderson). I.K. and K.B. acknowledge additional support from the Deutsche Forschungsgemeinschaft (DFG), SFB 951 (Project No. 182087777). Also, I.K. acknowledges financial support by the Stavros Niarchos Foundation (SNF) and the Hellenic Foundation for Research and Innovation (H.F.R.I.) under the 5th Call of “Science and Society” Action Always strive for excellence – ``Theodoros Papazoglou” (Project Number:11496, ``PSEUDOTOPPOS”). R.E. acknowledges support from the AFOSR Multidisciplinary University Research Initiative Award on Programmable Systems with Non-Hermitian Quantum Dynamics (Grant No.FA9550-21-1-0202), Army Research Office (W911NF-23-1-0312), and the Alexander von Humboldt Foundation. This collaboration was inspired by discussions during the Workshop ``Nonlinear Optics: Physics, Analysis, and Numerics" at Mathematisches Forschungsinstitut Oberwolfach. We are thankful to the institute for creating such a pleasant and stimulating atmosphere.

\end{acknowledgments}


\appendix

\section{Spectra and transport analysis of small arrays}
To illustrate the spectral and transport properties of small photonic arrays, we start with a pedagogical example of a $2\times2$ photonic system. The Hamiltonian for this system can be expressed as,

\begin{equation}
   H =  \begin{bmatrix}
        i\gamma & \kappa \\
        \kappa & -i\gamma
    \end{bmatrix}
\end{equation}
where $\gamma$ represents the gain/loss amplitude and $\kappa$ the coupling coefficient between the photonic elements. This simple system respects parity-time (\PT) symmetry, and when the gain/loss amplitude matches the coupling coefficient ($\gamma=\kappa$), it exhibits a second-order exceptional point. At this EP, both the eigenvalues and eigenvectors coalesce, indicating a critical phase transition and demonstrating the fundamental behavior of \PT-symmetric systems. 

\begin{figure*}
    \centering
    \includegraphics[width=0.95\textwidth]{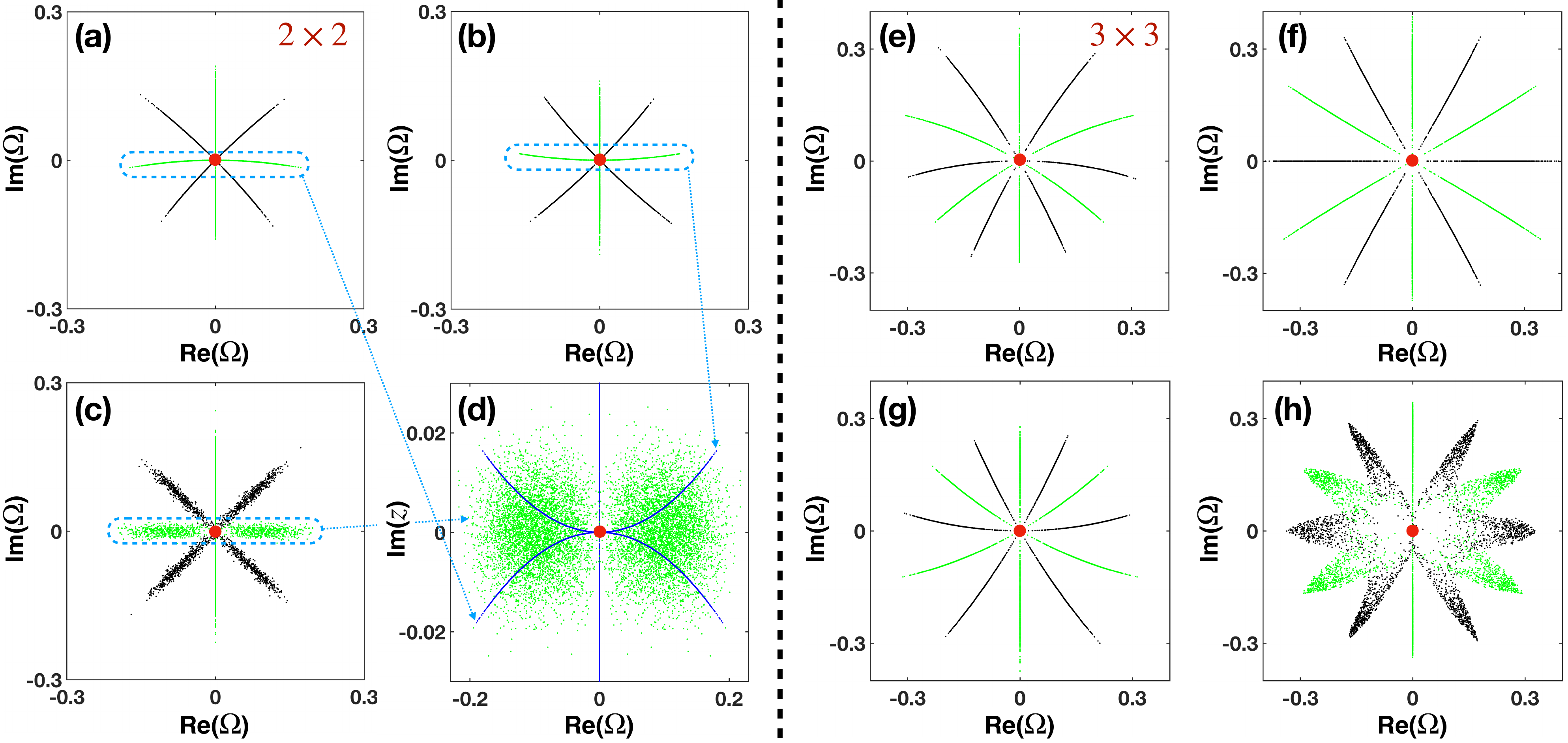}
    \caption{Demonstration of particle hole symmetry in (a)-(d) a $2\times2$ non-Hermitian system and (e)-(h) a $3\times3$ system. The perturbation strength is $\varepsilon = 0.01$ in all cases. Black dots correspond to real perturbations, while green dots represent purely imaginary ones. In (a) and (b), a single diagonal element of the Hamiltonian is perturbed, while in (c), both diagonal elements are perturbed, leading to the formation of extended cloud regions. (d) A zoomed-in view of the blue dashed lines further illustrates the cloud formation. In (e)-(g), only one diagonal element is perturbed, whereas in (h), all three diagonal elements are perturbed simultaneously, resulting in cloud regions when different curvature lines combine. Note that for purely imaginary perturbations, the distribution remains a straight line.}
    \label{fig:2x2_3x3}
\end{figure*}

In order to examine the spectral features of this system, we consider three different kinds of structured perturbations and analytically derive their eigenvalues,

\begin{equation}
    \mbox{Z}_1 = \begin{bmatrix}
        z_1 & 0 \\
        0 & 0
    \end{bmatrix}, \;\;
    \mbox{Z}_2 = \begin{bmatrix}
        0 & 0 \\
        0 & z_2
    \end{bmatrix}, \;\;
    \mbox{Z}_3 = \begin{bmatrix}
        z_1 & 0 \\
        0 & z_2
    \end{bmatrix}
\end{equation}
It is straightforward to show that at the EP ($\gamma = \kappa$) the eigenvalues of our perturbed Hamiltonian $H+\mbox{Z}_n$ with $n=1, 2, 3$ are,

\begin{align}
    &\Omega_j = \frac{z_1 \pm \sqrt{z_1^2 + 4iz_1}}{2} \nonumber \\
    &\Omega_j = \frac{z_2 \pm \sqrt{z_2^2 - 4iz_2}}{2} \nonumber \\
    &\Omega_j = \frac{z_1 + z_2 \pm \sqrt{(z_1 + z_2)^2 + 4i(z_1+z_2)}}{2}
\end{align}
respectively. In the above expressions we set $\gamma=\kappa=1$ (in normalized units) for simplicity and the indices $j=1,\;2$ denote the two eigenvalues. Since $z_j \ll 1$ we can simplify them with approximate relations as,

\begin{align}
    &\Omega_j \approx \frac{z_1}{2} \pm \sqrt{z_1}\sqrt{i} \nonumber \\
    &\Omega_j \approx \frac{z_2}{2} \pm 2\sqrt{z_2}\sqrt{-i} \nonumber \\
    &\Omega_j \approx \frac{z_1 + z_2}{2} \pm \sqrt{z_1+z_2}\sqrt{i}
\end{align}

Note here that the considered perturbations are either real ($z_j \in \mathbb{R}$) or purely imaginary ($z_j \in i\mathbb{R}$). In the case of real perturbations, the eigenvalues of $H+\mbox{Z}_1$ and $H+\mbox{Z}_2$ are always complex, indicating that the particle-hole symmetry (PHS) is not respected. However, when the perturbations are purely imaginary, we must examine two distinct scenarios: positive and negative perturbations. When the perturbations are positive, i.e., $z_j > 0$, the eigenvalues of $H+\mbox{Z}_1$ are purely imaginary, respecting the PHS, while the eigenvalues of $H+\mbox{Z}_2$ are complex. Conversely, when the perturbations are negative ($z_j < 0$), the behavior is reversed. 

Our results are shown in Figs.~\ref{fig:2x2_3x3}(a)-(d) on the complex plane, where black dots correspond to real perturbations and green dots represent purely imaginary perturbations. Fig.~\ref{fig:2x2_3x3}(a) shows the eigenvalue spectra of $H+\mbox{Z}_1$ for $l=1000$ realizations, Fig.~\ref{fig:2x2_3x3}(b) for $H+\mbox{Z}_2$, and finally, Fig.~\ref{fig:2x2_3x3}(c) for $H+\mbox{Z}_3$, with a perturbation strength of $\varepsilon = 0.01$ in all cases. Fig.~\ref{fig:2x2_3x3}(d) zooms into the blue dashed areas of each subplots, where the color in Figs.~\ref{fig:2x2_3x3}(a),(b) has been changed from green to blue for improved visualization and to avoid any potential confusion. When there is only one perturbation, the dependence remains a line, despite the breaking of PHS. However, when perturbations are applied to all elements, an elongated cloud forms in the regions where PHS is broken. In the former case, the dependence on perturbations is described by an invertible function, whereas in the latter, it is not. We emphasize here the formation of such clouds when the combined effect of perturbations arises from lines with different slopes. It is important to note that the pure imaginary line remains a straight line when both perturbations ($\mbox{Z}_3$) are considered. This is a consequence of PHS, as previously explained, since it is conserved in these cases. 

Additionally, we examined a $3\times3$ photonic system, as shown in Figs.~\ref{fig:2x2_3x3}(e)-(h). Similar to the $2\times2$ system, we applied a perturbation strength of $\varepsilon =  0.01$ and $l=1000$ realizations. Figs.~\ref{fig:2x2_3x3}(e)-(g) correspond to perturbations applied to a single element of the diagonal, while Fig.~\ref{fig:2x2_3x3}(h) represents the case where the full diagonal is perturbed. As observed previously, whenever we have lines with different slopes, an elongated cloud forms, while the pure imaginary straight line is preserved.

\begin{figure}
    \centering
    \includegraphics[width=0.47\textwidth]{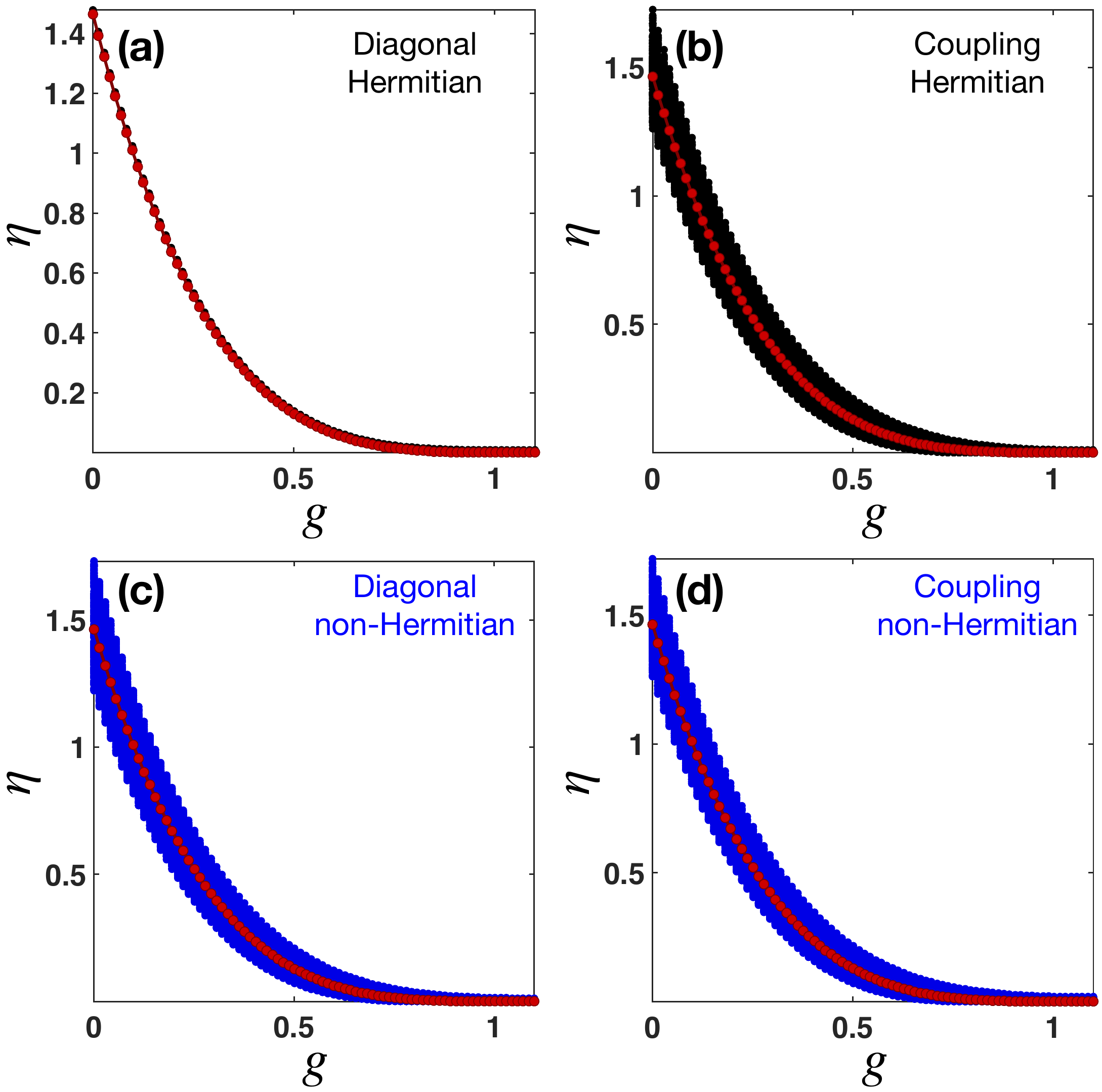}
    \caption{Wave transport dynamics for a $3\times3$ system under different types of disorder. The red line represents the theoretical relation. (a), (b) Hermitian disorder (black dots) applied to the diagonal and off-diagonal elements, respectively. (c), (d) The same analysis for non-Hermitian disorder (blue dots).}
    \label{fig:Transport_3x3}
\end{figure}

Having understood the behavior of the eigenvalues in the complex plane for both Hermitian and non-Hermitian perturbations, we now study the transport properties of the system. We consider a configuration consisting of either two or three coupled cavities, each subjected to the same uniform gain. In this scenario, we can derive analytical expressions for the transport $\eta$ both in the presence and absence of disorder. The dynamics of the system are governed by Eq.~\ref{eq:CMT}, with

\begin{align}
    &H_{c,2\times2} = \begin{bmatrix}
        \omega_0 - i\delta &\kappa \\
        \kappa & \omega_0 - i\delta
    \end{bmatrix} \nonumber \\
    &H_{c,3\times3} = \begin{bmatrix}
        \omega_0 - i\delta &\kappa & 0 \\
        \kappa & \omega_0 + ig & \kappa \\
        0 & \kappa & \omega_0 - i\delta
    \end{bmatrix}
\end{align}
where $\omega_0$ is the bare resonant frequency of the degenerate modes of the cavities in the absence of coupling. Parameter $\delta = \gamma_w - g$ is a non-Hermitian term that accounts for material gain $g$ and radiation loss through the waveguides $\gamma_w$. All the cavities consist of lossy materials, so an additional lossy term, denoted as $-i\gamma_c$, can be added in all the diagonal elements. However, in the subsequent analysis, we omit this term without loss of generality, as it can be removed through a simple gauge transformation. 

When excitation is applied from one of the ports, the steady-state field components and the asymmetric power distribution between modes (transport) have been calculated and are given as,

\begin{equation}
    \eta_2 = \frac{\delta^2}{\kappa^2}\;,\;\;\; \eta_3 = \frac{(\kappa^2 - \delta g)^2}{\kappa^4}
    \label{eq:Transport}
\end{equation}
where the subscript indices indicate the size of the system, as shown in previous studies \cite{Uniform}. It is evident that the strength of the uniform gain parameter, $g$, significantly influences the transport dynamics between the cavities. Unlike an array of waveguides, where a uniform gain or loss distribution typically results in power amplification or dissipation, we observe a completely different outcome in this setup. Here, the uniform gain not only enhances the system's response but also fundamentally alters the optical energy distribution within the structure. This behavior is not dependent on the engineering of the refractive index profile or a specific set of design parameters; rather, it is consistently observed across different configurations, regardless of the chosen design parameters and uniform gain distribution. 

We now introduce disorder into the system and examine two scenarios: diagonal disorder and off-diagonal (or coupling) disorder. The corresponding matrices for the $2\times2$ and $3\times3$ systems are,

\begin{align}
    &\mbox{Z}_d = \begin{bmatrix}
        z_1 & 0 \\
        0 & z_2
    \end{bmatrix}\;,\;\;\; 
    \mbox{Z}_c = \begin{bmatrix}
        0 & z_1 \\
        z_1 & 0
    \end{bmatrix}\;,\;\;\; \nonumber \\
    &\mbox{Z}_d = \begin{bmatrix}
        z_1 & 0 & 0 \\
        0 & z_2 & 0 \\
        0 & 0 & z_3
    \end{bmatrix}\;,\;\;\; 
    \mbox{Z}_c = \begin{bmatrix}
        0 & z_1 & 0 \\
        z_1 & 0 & z_2 \\
        0 & z_2 & 0
    \end{bmatrix}
\end{align}
respectively, where $``d"$ denotes diagonal and $``c"$ refers to coupling. By recalculating the transport properties for these modified Hamiltonians and modes, we observe that the transport behavior varies depending on the nature and strength of the disorder as follows,

\begin{align}
    &\eta_{2,d} = \frac{\delta^2 + z_2 ^2}{\kappa^2} \;,\;\;\; \eta_{2,c} = \frac{\delta^2}{(\kappa + z_1)^2} \nonumber \\
    &\eta_{3,d} = \frac{(\kappa^2 - \delta g -z_2z_3)^2 + (z_2\delta - z_3g)^2}{\kappa^4} \nonumber \\
    &\eta_{3,c} = \frac{(\kappa^2-\delta g + 2z_2\kappa + z_2^2)^2}{(\kappa+z_1)^2(\kappa+z_2)^2}
    \label{eq:Transport_P}
\end{align}

Our results for the transport behaviour of a $3\times3$ non-Hermitian disordered array of cavities are shown in Fig.~\ref{fig:Transport_3x3}, where we plot the transport as a function of the uniform gain for the same values of the parameters as in the main text. It is important to highlight that our system operates below the lasing threshold which is at $g\approx 1.11$. In all subplots, the red line represents Eq.~\ref{eq:Transport}, and the disorder strength is of the order of $W = 0.01$ for $l=1000$ realizations for each value of $g$. Notably, the average of all realizations yields the red line. Furthermore, the impact of diagonal non-Hermitian disorder is found to be greater than that of Hermitian disorder, as indicated by Eq.~\ref{eq:Transport_P} when $z_n \in i{\rm I\!R}$. Our results remain quantitatively consistent for both weaker and stronger disorder.

\section{Transport in large arrays}

\begin{figure}
    \centering
    \includegraphics[width=0.47\textwidth]{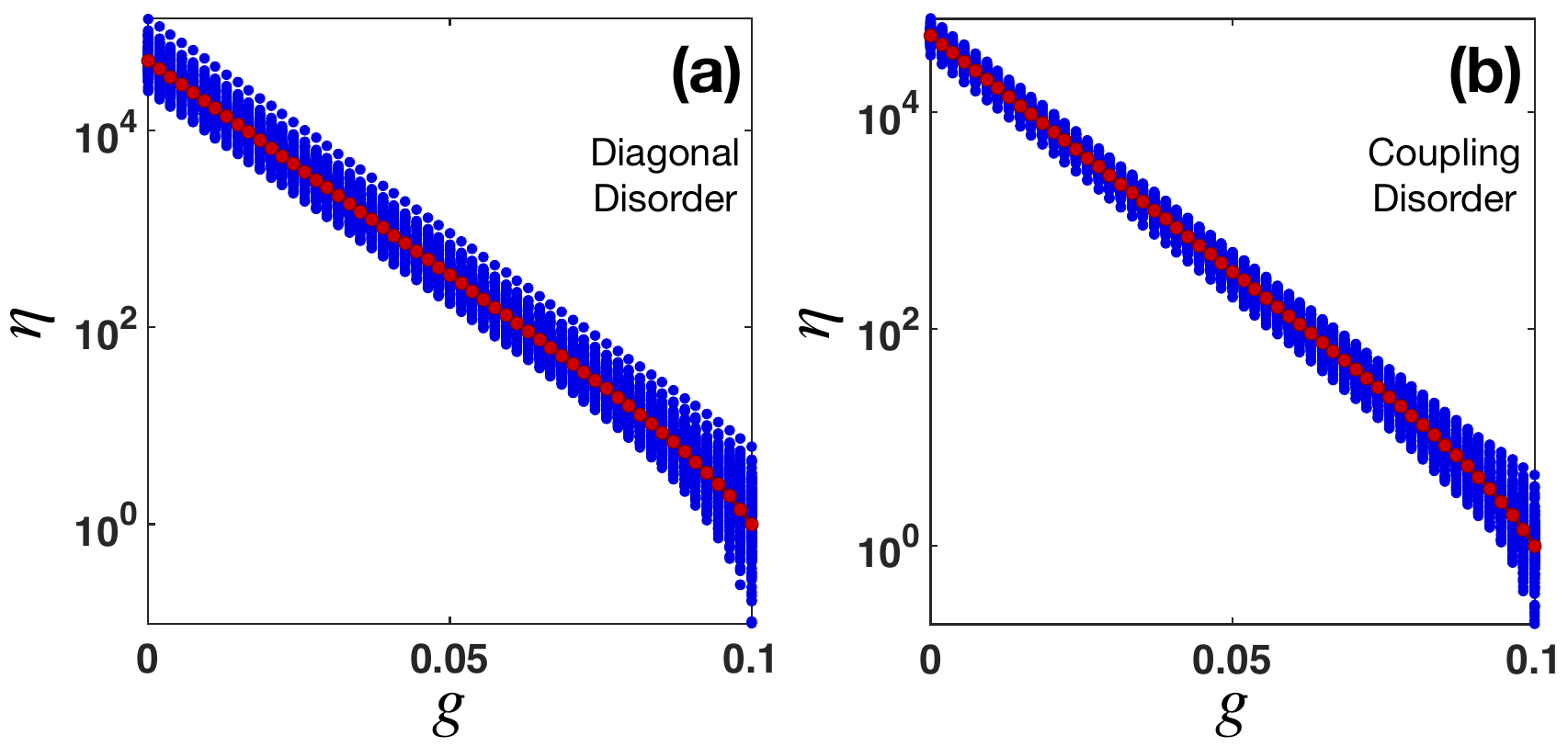}
    \caption{Wave transport on a logarithmic scale for a system of $N=101$ coupled cavities under (a) diagonal disorder and (b) off-diagonal disorder of strength $W=0.1$. Even for such large arrays, the behavior is consistent with the results presented in the main text.}
    \label{fig:Transport_101}
\end{figure}

\begin{figure}
    \centering
    \includegraphics[width=0.47\textwidth]{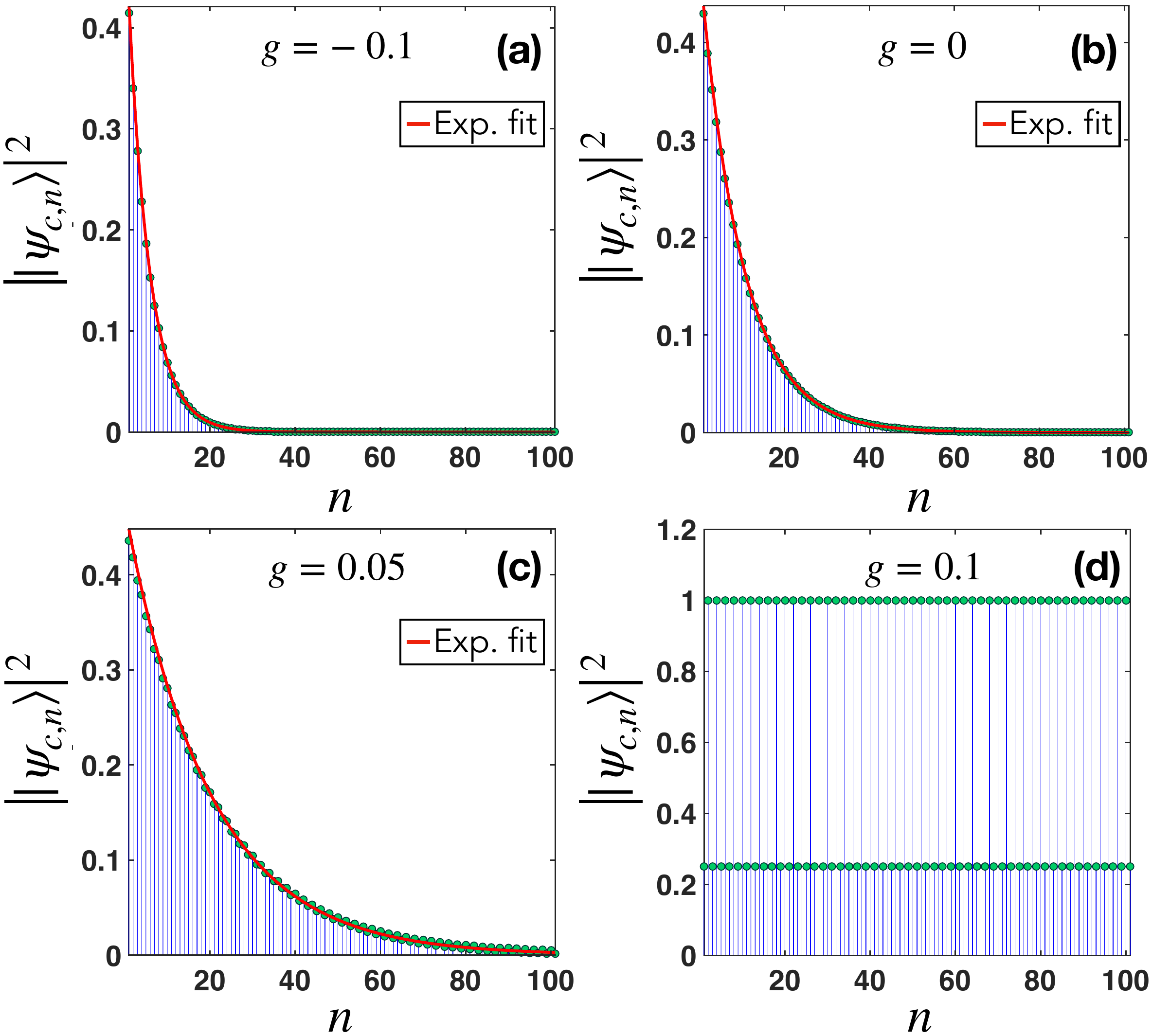}
    \caption{Intensity distribution across each cavity for a system with $N=101$ cavities under varying levels of uniform gain, $g$ and no disorder. (a) $g=-0.1$ corresponding to higher cavity losses, (b) $g=0$, (c) $g=0.05$, and (d) $g=0.1$. Note that the exponential decay (red line) persists even at higher values of $g$ than in the case of $N=41$ cavities, until just before the lasing threshold, where the optical power becomes evenly distributed throughout the lattice.}
    \label{fig:Power_101}
\end{figure}

In this appendix, we analyze a large array consisting of $N=101$ coupled cavities, using the same parameter values as outlined in the main text. We calculate the transport properties under both diagonal and off-diagonal disorder, with a perturbation strength of $W = 0.1$. Additionally, we examine the intensity distribution within the cavities for varying levels of uniform gain, $g$. This analysis extends the insights gained from smaller arrays, allowing us to explore the behavior of the system on a much larger scale. 

Figures \ref{fig:Transport_101}(a),(b) show the transport $\eta$, on a logarithmic scale as a function of the uniform gain $g$ for both diagonal and off-diagonal disorder, respectively. The results align with those observed in smaller arrays, showing that as the system approaches the lasing threshold, the transport value converges to one. It is important to note that the high values of $\eta$ observed when $g = 0$ are expected due to the highly localized modes at the edges of the array, which are a consequence of parameter $\gamma_w$. 

Finally, in Fig.~\ref{fig:Power_101} we compute the intensity distribution using the same parameters as in Fig.~\ref{fig:Power}. For larger arrays, we observe that the exponential decay persists even for higher values of uniform gain, until the system eventually reaches a steady-state distribution. This behavior indicates that the effects of gain on the intensity profile are consistent across varying array sizes.


\end{document}